\documentclass[12pt]{JINST}

\usepackage{graphicx,wrapfig}
\usepackage{mathptmx}           

\def\bec{\begin{center}}
\def\eec{\end{center}}
\def\beq{\begin{equation}}
\def\eeq{\end{equation}}
\def\benu{\begin{enumerate}}
\def\eenu{\end{enumerate}}
\def\bit{\begin{itemize}}
\def\eit{\end{itemize}}
\def\beqr{\begin{eqnarray}}
\def\eeqr{\end{eqnarray}}
\def\beqrs{\begin{eqnarray*}}
\def\eeqrs{\end{eqnarray*}}
\def\btab{\begin{tabbing}}
\def\etab{\end{tabbing}}
\def\btable{\begin{tabular}}
\def\etable{\end{tabular}}
\def\rarw{\rightarrow}
\def\Rarw{\Rightarrow}

\def\gm{\gamma}

\def\lm{\lambda}
\def\eps{\epsilon}

\def\Dl{\Delta}
\def\sg{\sigma}

\def\rarw{\rightarrow}

\def\del{\partial}

\def\bt{\beta}

\def\lan{\langle}
\def\ran{\rangle}

\title{Diffusion dynamics in a Tevatron store}

\author{Tanaji Sen \\ Fermilab, PO Box 500, Batavia, IL 60510 \\
Email: \email{tsen@fnal.gov}}

\abstract{A separator failure during a store in 2002 led to a
drop in luminosity, to increased emittance growth and to a drop in
beam lifetimes. We show that a simple diffusion model can be used
to explain the changes in emittance growth and beam lifetimes.}

\begin{document}

\section{Introduction}

Emittance growth of beams when they are in collision occurs due to 
many sources: beam-beam interactions, magnetic nonlinearities, 
intra-beam scattering, scattering off the residual gas and possibly
others. The dynamics of the emittance growth is complicated and it
depends strongly on the tunes. It is not always clear that the
dynamics can be described by a diffusion process at all particle
amplitudes in each beam. However in one store early in Run II, there
was a sudden drop in a separator voltage in the Tevatron and the
subsequent enhanced emittance growth and intensity lifetime drop
could be described by a simple diffusion model. 
In this report we analyze the luminosity drop, compare the
measured value with the expected drop and analyze the change in beam 
lifetimes. We show how a simple model of diffusive
emittance growth and a change in physical aperture provides a quantitative
explanation for the change in lifetimes.

\section{Separator failure and luminosity drop}

After about 13.5 hours into store 1253 (April 26, 2002), the voltage on 
the bottom plate of the horizontal
separator at A49 dropped from a value of -90kV to -25kV. This immediately
lowered the luminosity at CDF and D0. The emittance growth rate increased
and the lifetimes of both protons and anti-protons fell.
Table \ref{table: param} shows some of the key beam and machine 
parameters. 

\begin{table}
\bec
\btable{|l|c|} \hline
$\bt_x^*$ at CDF, D0 &   0.35m  \\
Horizontal tune & $\nu_x$ = 20.585 \\
Observed Luminosity drop at CDF &  41.4\% \\
Observed Luminosity drop at D0 &  42.3\% \\
Total proton beam intensity before drop & 5.78 $\times 10^{12}$ \\
Initial emittances (p and pbar) $\eps_x, \eps_y$ [$\pi$mm-mrad] & 22, 21 \\
Estimated final emittances $(p,\bar{p})$ [$\pi$mm-mrad] & (26 - 30, 25 - 29) \\
Average emitt. growth rate [$\pi$mm-mrad/hr] & 0.3-0.5 \\
Length of store [hrs] & 15 \\
Location of BPMs nearest to CDF [m] & 7.5 upstream and downstream \\ 
BPM upstream of CDF & $\bt_x$= 159.5 m, $\psi_x = 2\pi\times$ 20.337 \\
BPM downstream of CDF & $\bt_x$= 160.44 m, $\psi_x = 2\pi\times$ 0.238 \\
A49H Separator & $\bt_x$= 867.67m, $\psi_x = 2\pi\times$ 20.329 \\
\hline
\etable
\eec
\caption{Relevant beam and machine parameters in Store 1253.}
\label{table: param}
\end{table}

\begin{figure}
\centerline
{\includegraphics[scale=0.8]{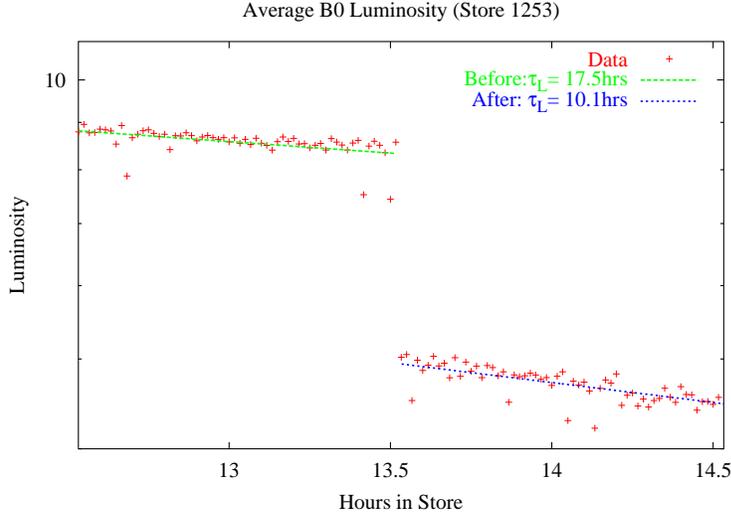}}
\caption{Average CDF Luminosity (raw data) and fits before and after the separator failure. The luminosity at CDF dropped by 41\% right after the separator failure.}
\label{fig: av_B0_lumi}
\end{figure}
Figure \ref{fig: av_B0_lumi} shows the measured luminosity drop at 
the CDF experiment after the separator failure. The luminosity at
D0 dropped by a similar amount. 

\subsection{Closed orbit shift at CDF and D0}

Here we estimate the orbit change as a consequence of the change  in
separator voltage at A49. 
The shift in the closed orbit due to a kick $\Dl\theta$ can be found from
\beq
\Dl x(s) = \frac{\Dl\theta\sqrt{\bt_{sep}\bt(s)}}{2\sin\pi\nu_x} 
\cos[\pi\nu_x - |\psi(s)-\psi_{sep}|]
\eeq
where $\bt_{sep}, \bt(s)$ are the beta functions at the separator and at $s$
respectively, $\psi(s),\psi_{sep}$ are the phase advances from a reference
point to $s$ and the separator respectively and $\nu_x$ is the horizontal tune.
The kick resulting from the electric field ${\cal E}$ across the separator plates of length $L$ on a particle of energy $E$ is given approximately by 
$\Dl \theta = {\cal E}L/E$.
At 980 GeV, a change in voltage of 65 kV across the separator plates with a gap of 5cm
results in a kick of
\beq
\Dl\theta = 3.41 \mu{\rm rad}   \label{eq: kick_sep}
\eeq

Using the above expressions we find that the proton's horizontal closed orbit would move by
\beq
\Dl x_{co}(CDF)|_{sep} =  29.9 \mu{\rm m}, \;\;\;\;\;\;\;\;  
\Dl x_{co}(D0)|_{sep} =  -30.3 \mu{\rm m}   \label{eq: xco_sep}
\eeq
The orbit separation $d_{co}$ between the beams at the IPs would be twice the above value if
the protons and anti-protons undergo the same but opposite changes in orbit. However the beam-beam
kick with separated beams also induces an orbit shift and this will be larger for the 
anti-protons. Calculation of the beam-beam induced orbit kick requires that we know the separation
between the beams, but that is precisely the quantity that we want to predict. 
We will 
approximate the beam-beam induced orbit kick by assuming that the beam separation is
twice the shift in the proton orbit. In that case, the beam-beam kick 
assuming round Gaussian beams is
\beq
\Dl x'_{bmbm} = 8\pi\xi\eps \frac{x+d}{(x+d)^2 + y^2}
\left\{ 1 - \exp[-\frac{(x+d)^2 + y^2}{2\sg^2}] \right\}
\eeq
where $\xi$ is the beam-beam parameter, $\eps$ is the un-normalized rms
emittance and $d$ is the horizontal separation between the beams.
Extracting the dipole part of the kick,
\beq
\Dl x'_{bmbm}(0,0) = 8\pi\xi\eps\frac{1}{d} \left\{ 1 - \exp[-\frac{d^2}{2\sg^2}]\right\}
\eeq
Since the sign of the pbar orbit offsets at CDF and D0 due to the separators 
have opposite signs,
the kicks experienced by the anti-protons due to the dipole beam-beam kicks at CDF and D0
have opposite signs. Hence
the contribution of the beam-beam kicks at CDF and D0 to the orbit shift at CDF is
\beq
\Dl x_{co}(CDF)|_{BB} = \frac{\bt^* |\Dl x'_{bmbm}|}{2\sin\pi\nu_x}
\left[ \cos\pi\nu_x - \cos(\pi\nu_x - |\psi_x(D0) - \psi_x(CDF)|) \right]  \label{eq: xco_bb}
\eeq
We have assumed here that the beta functions at the IPs did not change much.

From the average proton bunch intensity of $N_p = 1.61\times 10^{11}$ and an expected
proton emittance of 30$\pi$ at this stage of the store (this number is found later by
a self-consistent calculation), we find that the beam-beam parameter for anti-protons
at this stage was $\xi = 3.92\times 10^{-3}$. 
Hence the beam-beam kick using the value of the orbit offset found in 
Equation (\ref{eq: xco_sep}) is
\beq
\Dl x'_{bmbm}(CDF) = 5.14 \mu{\rm rad}
\eeq
while at D0, the kick has the opposite sign. Note that this kick is larger than the
kick due to the change in the separator voltage, cf. Eq. (\ref{eq: kick_sep}). However
because of the small beta function at the IPs, the change in orbit due to these beam-beam
kicks is quite small,
\beq
\Dl x_{co}(CDF)|_{BB} = -0.42 \mu{\rm m}
\eeq
using Equation (\ref{eq: xco_bb}). This is almost two orders of magnitude smaller than
the orbit shift due to the separator and can be neglected.

The predicted luminosity in terms of the luminosity ${\cal L}_0$ before the separator
failure is found from
\beq
{\cal L} = {\cal L}_0 \exp[- \frac{d_{co}^2}{2(\sg_p^2 + \sg_{\bar{p}}^2)}]
\eeq
where $d_{co} \approx 2\Dl x_{co}|_{sep}$. This calculation depends on the emittances
at the time of the failure. With the initial emittances and average emittance growth rates
shown in Table 1, the proton emittances after 13.5 hrs were likely to be in the range 
(26, 30)$\pi$mm-mrad and anti-proton emittances (25, 29)$\pi$mm-mrad.
We find using the orbit shifts in Eq. (\ref{eq: xco_sep}) (virtually the same at CDF and D0)
and the low and high end of the emittance range,
\beqr
\frac{\Dl {\cal L}}{\cal L} =  0.47, \;\;\;\;  (\eps_p = 26, \eps_{pbar} = 25)
\nonumber \\
\frac{\Dl {\cal L}}{\cal L} = 0.42 \;\;\;\;  (\eps_p = 30, \eps_{pbar} = 29)
\eeqr
These values are to be compared with the observed relative drops in luminosity of 0.414
at CDF and 0.423 at D0. This suggests that the emittances were more likely at the higher
end of the quoted range.

Another test of the optics
is to propagate the measured orbit changes at the BPMs closest to CDF back to CDF using
\beq
\Dl x(s_2) = \sqrt{\frac{\bt(s_2)}{\bt(s_1)}}
\frac{\cos[\pi\nu - |\psi(s_2)-\psi_{sep}|]}{\cos[\pi\nu - |\psi(s_1)-\psi_{sep}|]} \Dl x(s_1)
\eeq
where $s_1$ is the location of a BPM and $s_2$ is the location of an IP.
This expression does not depend on the kick angle at the separator nor upon the 
beta function at the separator. 

\begin{figure}
\centering
\includegraphics[scale=0.55]{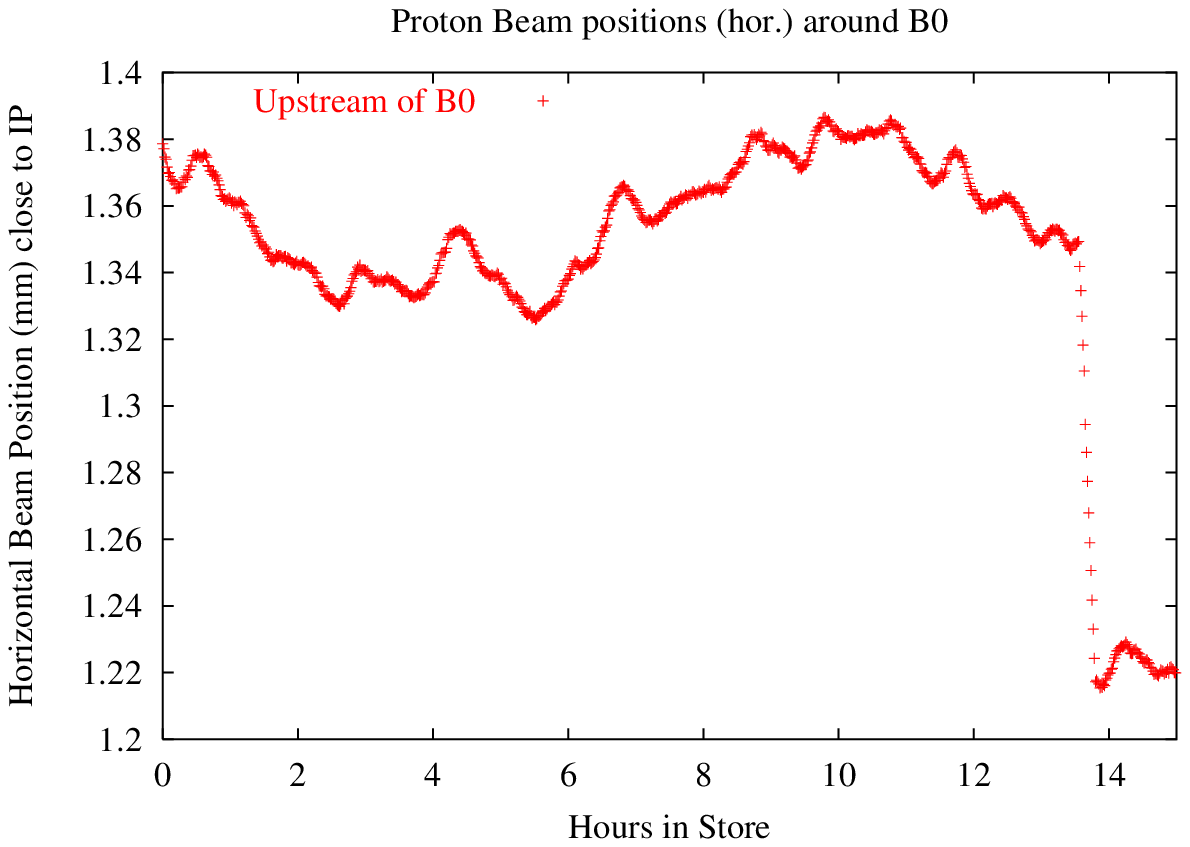}
\includegraphics[scale=0.55]{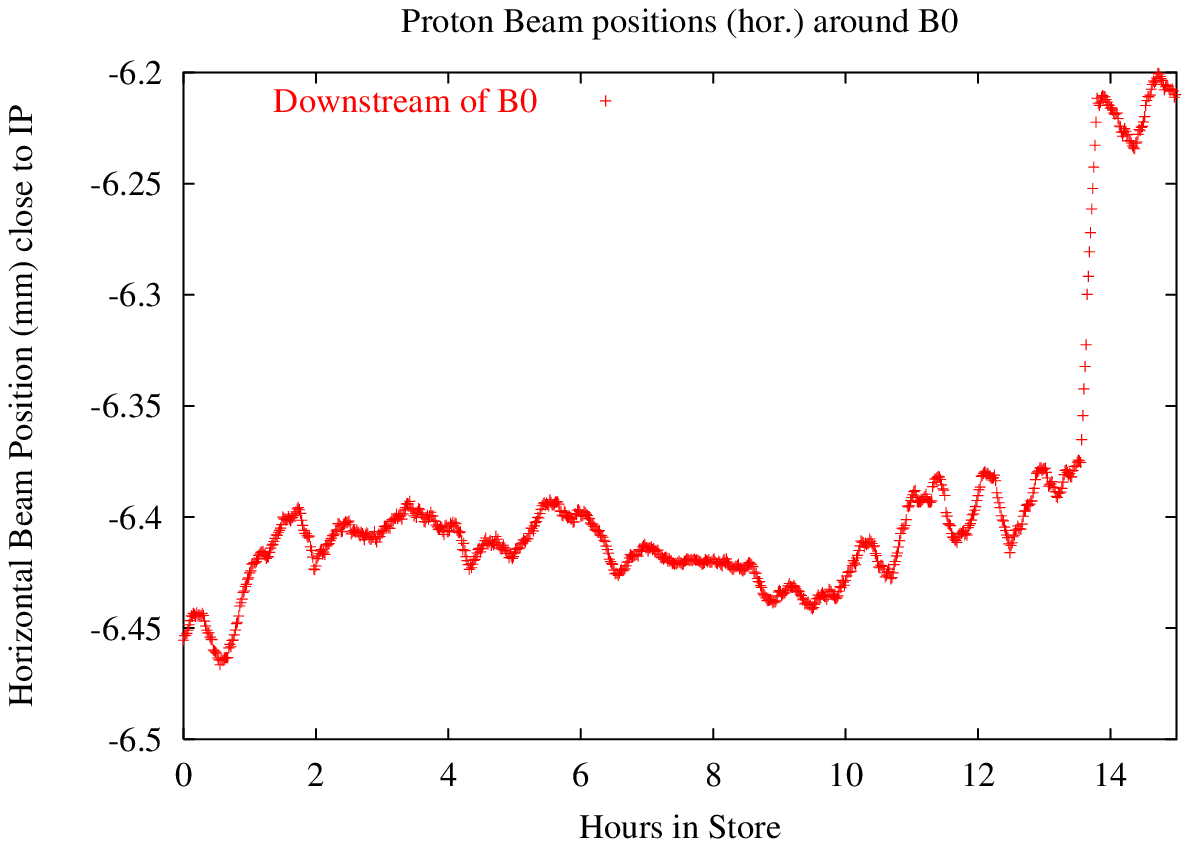}
\caption{Proton horizontal beam position at BPMs upstream and downstream of CDF. The jump
in position coincides with the separator failure}
\label{fig: pbeam_B0}
\end{figure}
From Figure \ref{fig: pbeam_B0} we observe that just before the failure, the proton
horizontal position was relatively steady at 1.349 mm, then it falls for about 15 minutes
after which it stabilized at 
1.217 mm. Similarly at the downstream BPM, the horizontal positions at these same times are
-6.375 mm and -6.214 mm. This slow decay in the position is related to the long 
integration time (about 15 mins) of these collision point monitors(CPMs).

The observed shifts at the CPMs around CDF were
\beq
{\rm Upstream}: \;\;\; \Dl x_U^{obs}  =   -0.132\; {\rm mm} 
\;\;\;\;\;\;\;\;\;\;\;\;\;\;\;\;\;\; 
{\rm Downstream}: \;\;\; \Dl x_D^{obs}  =   0.161\; {\rm mm} 
\eeq
Propagating these orbit shifts to the IP, we find using the upstream BPM that the expected
orbit shift at CDF and relative luminosity drops are
\beq
\Dl x_{co} (CDF) =  0.0285\; {\rm mm} , \;\;\;\;\;\; 
\Rarw \frac{\Dl {\cal L}}{\cal L}(CDF) = 0.384
\eeq
while using the downstream BPM, we find
\beq
\Dl x_{co} (CDF) = 0.0241\; {\rm mm} , \;\;\;\;\;\; 
\Rarw \frac{\Dl {\cal L}}{\cal L}(CDF) = 0.293
\eeq
This calculation shows that the upstream BPM was more consistent with the observed
luminosity drop. This may indicate either more errors downstream from the IP to the
CPM (this is unlikely since there is only the detector between the IP to the CPM) or 
that this downstream CPM reading was less reliable.

\section{Lifetimes and emittance growth times}

The luminosity in terms of beam parameters is
\beq
{\cal L} = \frac{3\gm f_{rev}M_b N_p N_{\bar{p}}}{\pi\bt^*(\eps_{N,p}+\eps_{N,\bar{p}})} 
{\cal H}(\frac{\bt^*}{\sg_s})
\eeq
where $f_{rev}$ is the revolution frequency, $M_b$ is the number of bunches in 
each beam, $N_p, N_{\bar{p}}$ are the proton and anti-proton bunch intensities
respectively, $\eps_{N,p},\eps_{N,\bar{p}}$ are the 95\% emittances of the 
beams, $\sg_s$ is the rms bunch length and ${\cal H}$ is
the hourglass form factor
\beq
{\cal H}(z) = \sqrt{\pi} z e^{z^2} (1 - \Phi(z)), \;\;\;\;\;\;\; 
\Phi(z) = \frac{2}{\sqrt{\pi}}\int_0^z e^{-t^2} dt
\eeq
The average bunch length in Store 1253 was 2.6 nano-seconds or $\sg_s=78$cm. With
$\beta^*=35$cm, the hourglass reduction factor for $z \equiv  \bt^*/\sg_s = 0.45$ is 
${\cal H}(z) = 0.51$. 

The luminosity lifetime can be calculated from the beam parameters by
taking the logarithmic derivatives.
Defining the luminosity and intensity lifetimes as
\beq
\frac{1}{\tau_{\cal L}} = - \frac{1}{\cal L} \frac{d{\cal L}}{dt}, \;\;\;\;\;\;\;\;\;
\frac{1}{\tau_p} = - \frac{1}{N_p} \frac{dN_p}{dt}, \;\;\;\;\;\;\;
\frac{1}{\tau_{\bar{p}}} = - \frac{1}{N_{\bar{p}}} \frac{dN_{\bar{p}}}{dt}
\eeq
while the longitudinal bunch length and transverse emittance growth times are
\beq
\frac{1}{\tau_s} = \frac{1}{\sg_s} \frac{d\sg_s}{dt}, \;\;\;\;\;\;\;\;\;
\frac{1}{\tau_{\eps_{\perp}}} = \frac{1}{\eps_{N,p}+\eps_{N,\bar{p}}}
\frac{d}{dt}(\eps_{N,p}+\eps_{N,\bar{p}})
\eeq
Then the luminosity lifetime is 
\beq
\frac{1}{\tau_{\cal L}} = \frac{z}{\cal H} \frac{d{\cal H}}{dz} \frac{1}{\tau_s}
 + \frac{1}{\tau_p} + \frac{1}{\tau_{\bar{p}}} + \frac{1}{\tau_{\eps_{\perp}}}
\eeq
Here $z = \bt^*/\sg_s$. 
Using the expression for the hour-glass form factor ${\cal H}$, this can be 
rewritten as an expression for the emittance growth rate,
\beq
 \frac{1}{\tau_{\eps_{\perp}}} = \frac{1}{\tau_{\cal L}} - 
(1 - \frac{2z}{\cal H} + 2z^2) \frac{1}{\tau_s}
 - \frac{1}{\tau_p} - \frac{1}{\tau_{\bar{p}}} 
\label{eq: tau_emitt}
\eeq
We calculate the emittance growth time $\tau_{\eps_{\perp}}$
from the measured values of the other time scales. This is useful because
at this stage in Run II the synchrotron light
monitor was not available, so there was no direct measurement of the
transverse emittance growth rate during the store. 

\begin{figure}
\centering
\includegraphics[scale=0.55]{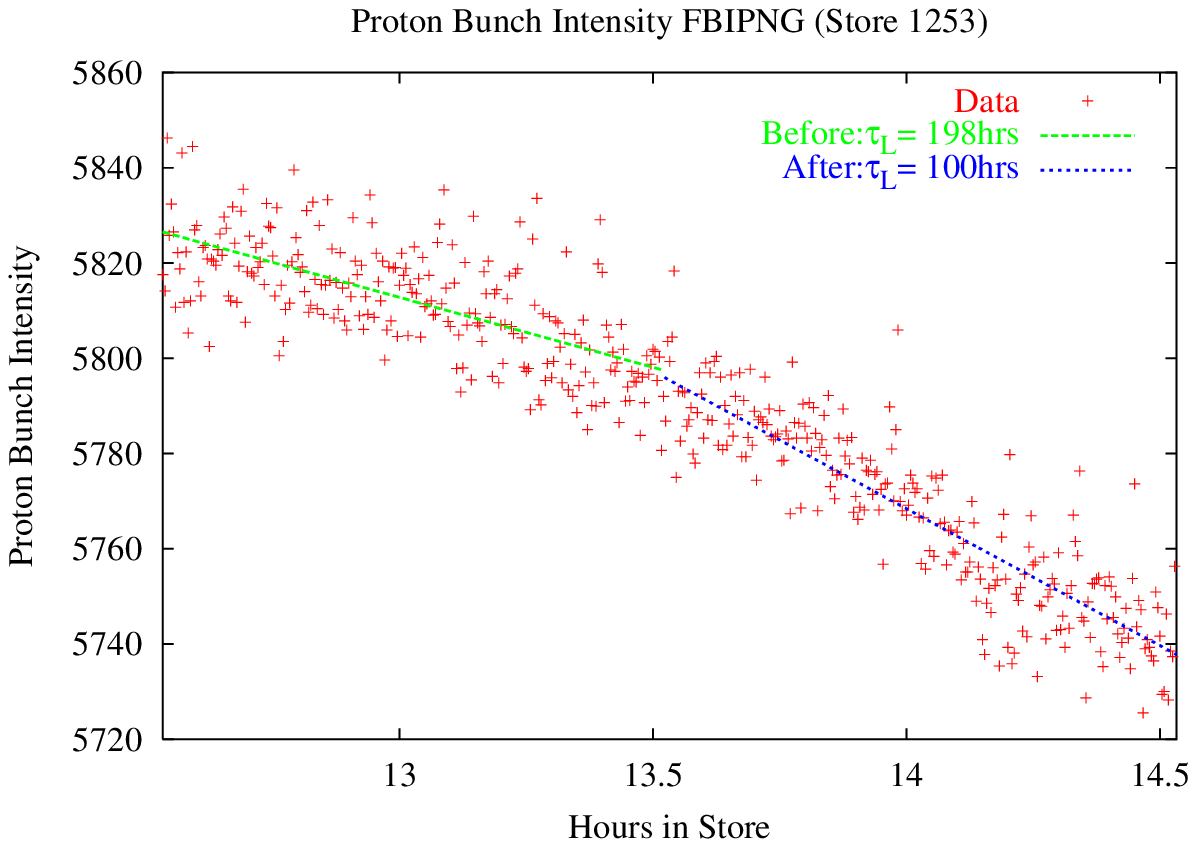}
\includegraphics[scale=0.55]{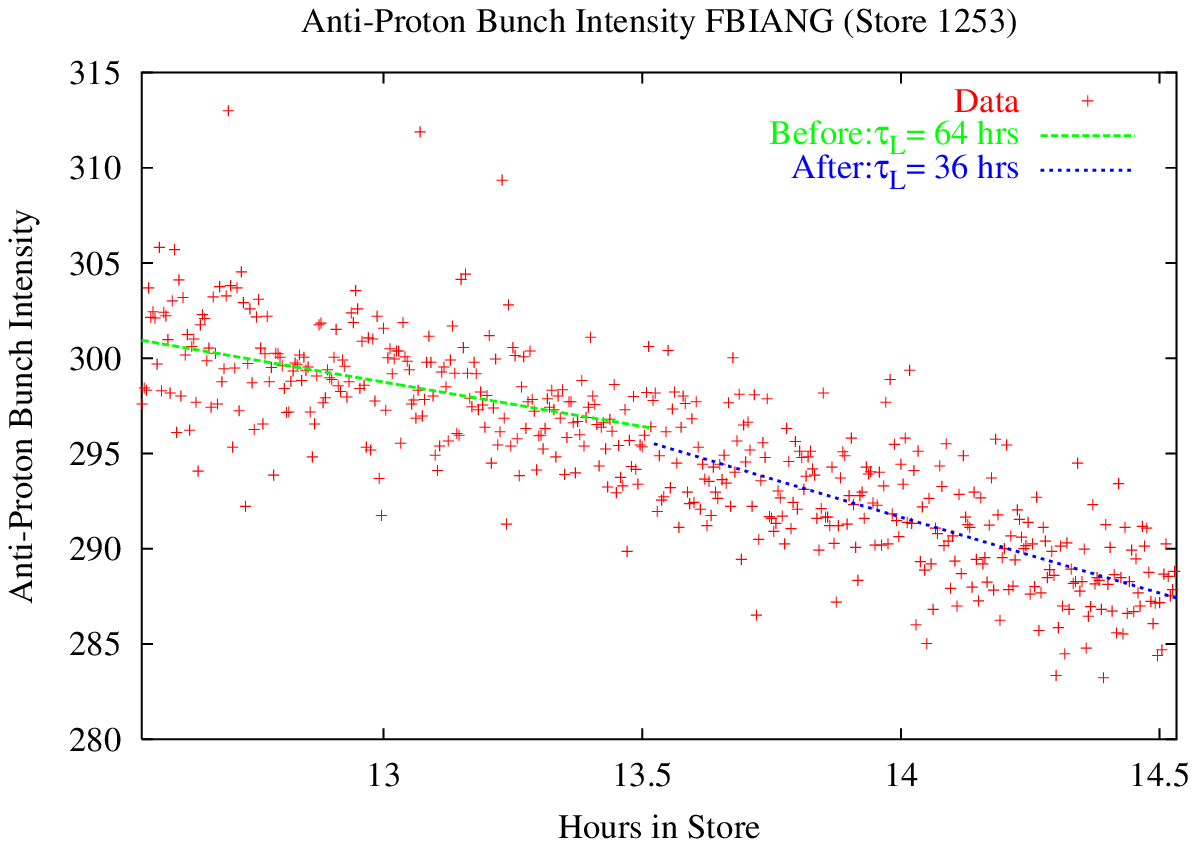}
\caption{Proton bunch intensity (left) and anti-proton bunch intensity (right) before and
after the separator failure. Note the change in slope at around 13.5 hours}
\label{fig: intensities}
\end{figure}
Figure \ref{fig: intensities} shows the proton and anti-proton bunch intensities an hour
before and after the separator failure. There is a clear change in the intensity lifetimes
before and after the failure.

\begin{figure}
\centering{\includegraphics[scale=0.6]{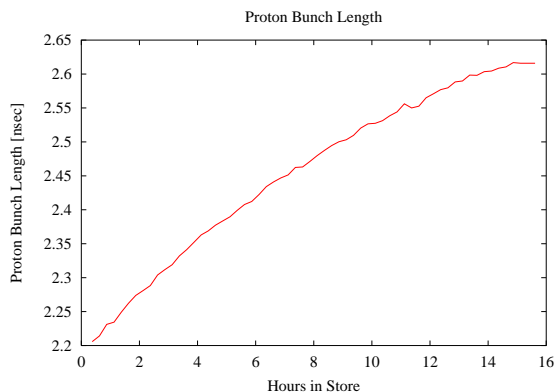}}
\caption{Proton bunch length during the store. The separator failure occurred
about at 13.5 hours into the store.}
\label{fig: plength}
\end{figure}
Figure \ref{fig: plength} shows the bunch length as a function of time over the
store. There is evident growth in the bunch length over the 15 hours of the store
but from the data then available (bunch lengths every 15 minutes) it was not 
possible
to discern a change in the growth of the bunch length after the separator failure.
We assume that the growth rates of the bunch length were the same an hour before
and an hour after the failure.  Table \ref{table: times} shows
the measured lifetimes and growth times and the calculated
transverse emittance growth time before and after the failure. 

\begin{table}[htb]
\bec
\btable{|l|c|c|} \hline
   & Before separator failure & After separator failure \\ \hline
Luminosity lifetime [hrs] & 17 & 10 \\
Proton lifetime [hrs] & 198 & 100 \\
Anti-proton lifetime [hrs] & 64 & 36 \\
Average Bunch length [nsec] & 2.6 & 2.6 \\
Bunch length growth time [hrs] & 81.5 & 81.5 \\
Transverse Emittance growth time [hrs] & 24.5 & 15.3 \\
\hline 
\etable 
\eec
\caption{Lifetimes and growth times from data 1 hour before and an hour after the 
separator failure. The transverse emittance growth rates are calculated from Equation (3.6).} 
\label{table: times}
\end{table}

\subsection{Beam lifetime}

The separator failure changed the beam orbits around the ring. 
The changes in the proton orbit were calculated using a MAD optics file. 
The maximum horizontal orbit change was 1.8$\sg$ radially outwards while the rms
orbit change was 0.59$\sg$. In the vertical plane, the maximum orbit shift was
about 1.8$\sg$ downwards but the rms orbit change was smaller, 0.24$\sg$, as expected. An important consequence of the change in orbit is a change in physical
aperture. Since the collimators define the limiting aperture, movement towards
them would reduce the physical aperture.
\begin{figure}
\centering{\includegraphics[scale=0.46]{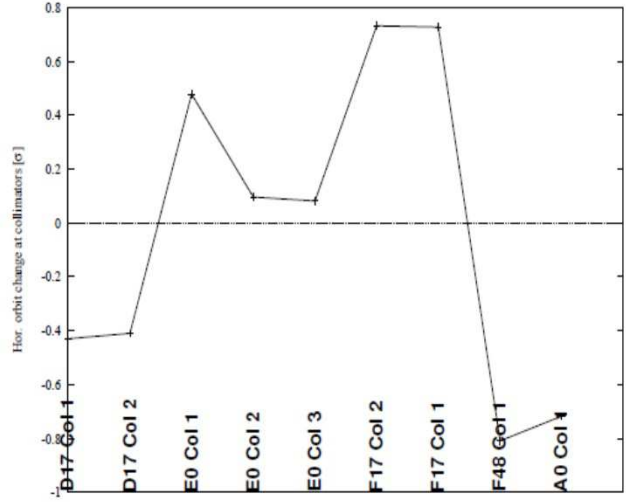}}
\caption{Change in the horizontal proton orbit (in units of the rms beam size) at the
collimators due to the separator failure. The largest change was at the F17 collimators
where the protons moved by about 0.73 $\sg$ to the radial outside.}
\label{fig: dxorbit_collim}
\end{figure}
Figure \ref{fig: dxorbit_collim} shows the simulated proton horizontal orbit 
change at several collimators in the ring. The largest change at a collimator
location was about 0.73$\sigma$ towards the F17 collimator.

We have seen  that the emittance growth rate increased
and the orbits changed significantly after the separator failure. Could these two
phenomena explain the sharp drop in beam lifetime?

We assume that the beam density distribution function evolves according
to the diffusion equation. For simplicity we will consider the transverse 
distribution function can be decoupled as the product of horizontal $\rho_x$ and 
vertical $\rho_y$ distribution functions and consider only the evolution of $\rho_x$,
\beq
\frac{\del}{\del t}\rho_x = \frac{\del}{\del J_x}\left[ D(J_x)
 \frac{\del\rho_x}{\del J_x} \right]
\eeq
Here $D(J_x)$ is the diffusion coefficient in the action 
$J_x = [x^2+(\bt_x x' + \alpha_x)^2]/\bt_x$. The limiting physical aperture is
assumed to be in the horizontal plane at an amplitude $A_x$ and corresponding
action at the aperture is $J_A = A_x^2/\bt_x$. Under the assumption of
independence of the transverse planes, the total number of particles in the beam
at time $t$ can be written as $N(t) = N_0 N_x(t) N_y(t)$ where $N_0$ is
the initial number of particles, $N_x$ is a scaled time dependent number 
defined by
\beq
N_x(t) = \int_0^{J_A} \rho_x(J_x, t) \; dJ_x, \;\;\; N_x(0) = 1
\eeq
and a similar expression for $N_y$. We assume that particles that diffuse out to 
the aperture $J_A$ are lost.

The beam emittance $\eps_x$ is related to the average action which is defined as
\beq
\lan J_x \ran = \int_0^{J_A} J_x \rho_C(J_x,t) \; dJ_x 
\eeq
where $\rho_C$ is the conditional density which accounts for the particle number
changing in time and hence is defined as $\rho_C = \rho_x/N_x$. From the definition
it is clear that $\rho_C$ is normalized to unity at all times.
It follows from the diffusion equation that
the average action evolves as 
\beq
\frac{d}{dt} \lan J_x \ran = \int_0^{J_A} D'(J_x) \rho_C \; d J_x + 
 [J_A - \lan J_x \ran] D(J_A) \frac{\del \rho_C}{\del J_x}(J_A)
\eeq
If the density falls sufficiently rapidly and the aperture is far enough away from
the beam, then $\del \rho_C(J_A)/\del J_x \rarw 0$ and the second term in the
above equation can be dropped. With this simplification,
\beq
\frac{d}{dt} \lan J_x\ran =  \int_0^{J_A} D'(J_x) \rho_C \; d J_x 
\eeq
If the diffusion coefficient increases linearly with the action,
\beq
D(J_x) = D_0 J_x    \label{eq: linear_dcoef}
\eeq
then it follows
\beq
\frac{d}{dt} \lan J_x\ran =  D_0
\eeq
The effective transverse emittance calculated from the luminosity was found to
increase nearly linearly with time during stores, we may therefore assume that 
Equation (\ref{eq: linear_dcoef}) is valid, at least in the core of the beam.

We can now calculate time scales associated with the diffusive motion. The 
{\em mean escape time} for particles to travel to the absorbing boundary at 
$J_A$ is defined as \cite{Gardiner}
\beq
t_{esc} = \int_0^{J_A} \frac{J_x}{D(J_x)} \; d J_x
\eeq
Assuming Equation (\ref{eq: linear_dcoef}), we obtain
\beq
t_{esc} = \frac{J_A}{D_0}   \label{eq: escapetime}
\eeq
This escape time is related to the beam lifetime.

The lifetime can be calculated by a more complete analysis as in Edwards and
Syphers \cite{EdSyph}. The diffusion equation can be solved analytically when
the diffusion coefficient is linear or quadratic in the action. With a linear
dependence as assumed above, the density at time $t$ is
\beq
\rho_x(J_x,t) = \sum_n c_n J_0(\lm_n\sqrt{\frac{J_x}{J_A}})
\exp[-\frac{\lm_n^2}{4} \frac{D_0 t}{J_A}]  \label{eq: rho_soln}
\eeq
where 
\beq
c_n = \frac{1}{J_1^2(\lm_n)J_A}\int_0^{J_A} \rho_0(\frac{J_x}{J_A})
J_0(\lm_n\sqrt{\frac{J_x}{J_A}}) \; d J_x
\eeq
$J_0, J_1$ are the zeroth and first order Bessel functions, the $\lm_n$'s are the
$n$'th roots of $J_0$ and $\rho_0$ is the initial density. For an initially
Gaussian distribution in phase space, the distribution in action is an exponential,
\beq
\rho_0(J_x) = \alpha \exp[-\frac{\alpha J_x}{J_A}] \; , \;\;\;\;\;\;
\alpha = \frac{A^2}{2\sg_x^2}
\eeq
Assuming that the beam is sufficiently far from the aperture, the coefficients
$c_n$ simplify in this case to 
\beq
c_n = \frac{1}{J_1^2(\lm_n)} \exp[-\frac{\lm_n^2}{4\alpha}]
\eeq
Keeping only the first and dominant term in the solution for the density 
Equation (\ref{eq: rho_soln}), the scaled partial number of particles $N_x$ in
the beam simplifies to
\beq
N_x(t) \simeq \frac{2}{\lm_1 J_1(\lm_1)} \exp[-\frac{\lm_1^2}{2}(\frac{\sg_x}{A})^2]
\exp[-\frac{\lm_1^2}{4} \frac{D_0 t}{J_A}]
\eeq
We assume that the lifetime was determined by particles reaching the
horizontal aperture first. 
It follows that the lifetime defined as
\beq
t_L = -\frac{N_x}{dN_x/dt} = \frac{4}{\lm_1^2}\frac{J_A}{D_0} \simeq 
0.7\frac{J_A}{D_0}  \label{eq: lifetime}
\eeq
This expression is very close to the mean escape time $t_{esc}$ calculated in 
Equation (\ref{eq: escapetime}). 

We will now use Equation (\ref{eq: lifetime}) to relate the beam lifetimes before and
after the separator failure. Equating
\beq
D_0 \equiv \frac{d\eps_x}{dt} = \frac{\eps_0}{\tau_{\eps}}
\eeq
where $\eps_0$ is the initial emittance and $\tau_{\eps}$ is the emittance growth
time. Then
\beq
t_L = 0.7 \frac{J_A}{\eps_0} \tau_{\eps}
\eeq
Before the separator failure, the beam aperture was approximately $6\sg$ at one
or more of the collimators. Hence before the failure, $J_A = (6\sg)^2/\bt_x$.
After the failure the beam moved closer to the physical aperture by 0.7$\sg$. Hence
$J_A = (5.3\sg)^2/\bt_x$ after the failure.
\beq
\Rightarrow \frac{t_L({\rm after})}{t_L({\rm before})} = 
\frac{J_A({\rm after})}{J_A({\rm before})} 
\frac{\tau_{\eps}({\rm after})}{\tau_{\eps}({\rm before})}
\eeq
From Table \ref{table: times} we find that $\tau_{\eps}({\rm before})=24.5$hrs
and $\tau_{\eps}({\rm after}) = 15.3$hrs. Hence
\beq
\frac{t_L({\rm after})}{t_L({\rm before})} = (\frac{5.3}{6})^2 \frac{15.3}{24.5}
 = 0.49
\eeq
From Table \ref{table: times} we find that the ratio of the measured 
lifetimes is 
$= 100/198 = 0.51$. Hence the predictions of the one dimensional theory are
in very good agreement with the observed drop in lifetime.
The increased emittance growth after the separator failure may have been due
to multiple sources resulting from the change in orbits including change in
tunes, increased long-range beam-beam effects from smaller separations at
some locations and larger nonlinear fields in some magnets etc.

No matter what the sources of emittance growth were, we have shown that the drop 
in lifetime after the separator failure was consistent with a simple model of
diffusive 
emittance growth and the beam center moving closer to a physical aperture.


\begin{thebibliography}{99}
\bibitem{EdSyph}D. Edwards and M.J. Syphers,{\em An Introduction to the Physics of
High Energy Accelerators}, John Wiley (1993)
\bibitem{Gardiner}C.W. Gardiner,{\em Handbook of Stochastic Methods},
 Springer-Verlag (1985)
\end{thebibliography}
\end{document}